\documentclass[a4paper,12pt]{article} 

\usepackage{geometry,epsfig,subfigure}
\usepackage{amsmath}
\usepackage{amssymb}
\usepackage{graphics,epsfig,mydef}

\usepackage[square,sort,comma,numbers]{natbib}

\usepackage[usenames]{color}
\usepackage{fancyhdr,url}   
\usepackage{setspace}  
\usepackage{multirow}   
\usepackage{rotating}    
\usepackage{colortbl}     
\usepackage{relsize} 
\usepackage{booktabs}
\usepackage{cancel}
\usepackage{enumerate}
\numberwithin{equation}{section}
\usepackage{soul} 
\usepackage{xcolor} 
\usepackage{pdflscape}
\usepackage{hyperref}
\usepackage[capitalise]{cleveref}
\usepackage{lineno}
\usepackage{graphicx}
\usepackage{etoolbox}

\definecolor{darkblue}{rgb}{0,0,0.8}
\definecolor{darkgreen}{rgb}{0,0.5,0}

\long\def\symbolfootnote[#1]#2{\begingroup \def\thefootnote{\fnsymbol{footnote}}\footnote[#1]{#2} \endgroup} 

\usepackage{amsthm}

\begin{document}
\renewcommand*{\thepage}{\arabic{page}}

\setstretch{1.3}

\begin{center}
\Large
\textbf{Molecular Diffusion Replaces Capillary Pumping in Phase-Change Driven Nanopumps\\}
\normalsize
\vspace{0.2cm}
Yigit Akkus$^{a,b}$, Ali Beskok$^a\symbolfootnote[1]{e-mail: \texttt{abeskok@smu.edu}}\!$ \\
\smaller
\vspace{0.2cm}
$^a$Lyle School of Engineering, Southern Methodist University, Dallas, Texas 75205, USA \\
$^b$ASELSAN Inc., 06172 Yenimahalle, Ankara, Turkey\\
\vspace{0.2cm}
\end{center}

\textbf{Nano-scale fluid transport has vast applications spanning from water desalination to biotechnology \cite{shannon2008,whitby2007}. It is possible to pump fluids in nano-conduits using pressure gradients \cite{koplik1988}, thermal methods \cite{longhurst2007}, electric \cite{qiao2003,huang2007} and magnetic fields \cite{li2013}, and with manipulations of surface chemistry and electric fields \cite{rinne2012,luca2013,zhang2016}. Inspired by the capillary-driven phase change heat transfer devices, we present a phase-change driven nanopump operating almost isothermally. Meticulous computational experiments on different sized nanopumps revealed efficient operation of the pump despite the reduction in system size that extinguishes capillary pumping by annihilating the liquid meniscus structures. Measuring the density distribution of liquid in cross sections near to the evaporating and condensing liquid-vapor interfaces, we discovered that phase change induced molecular scale mass diffusion mechanism replaces the capillary pumping in the absence of meniscus structures. Therefore, proposed pumps can serve as a part of both nanoelectromechanical (NEMS) and microelectromechanical systems (MEMS) with similar working efficiencies, and can be used for continuous gas separation applications.}

Nanotechnology enabled fabrication of solid state nanostructures and devices that interact with the surrounding liquid and gas environments \cite{iijima1991,qin2000}. With scale reduction, interfacial and surface forces dominate over body forces. Eventually, the fluid cannot be treated as a continuous media since its molecular nature and atomistic interactions become increasingly important in determining the transport phenomena \cite{karniadakis2006}. Thermal gradients are known used to induce continuous liquid flow through nano-conduits \cite{longhurst2007,liu2010,liu2012,thekkethala2013,zhao2015}, and transport liquid nanoclusters inside CNTs \cite{zambrano2009,shiomi2009}. Continuous flow of liquid Argon from low to high temperature reservoirs through CNTs and graphene channels was associated with thermal creep in density layers \cite{thekkethala2013}, while the flow direction was reversed in channels with very low surface energy \cite{liu2012}. Thermal creep phenomenon, which is also relevant for nano-scale gas transport, requires large temperature gradients \citep{karniadakis2006}. In this communication, we demonstrate phase change as the driving mechanism of a continuous flow nanopump, which operates at nearly isothermal conditions. The pump is regulated by simultaneous cooling and heating at the opposite ends of the channel (condenser and evaporator regions, respectively), and it works for as long as the liquid wets the wall.

\begin{figure}[t]
	\centering	
	\epsfig{file=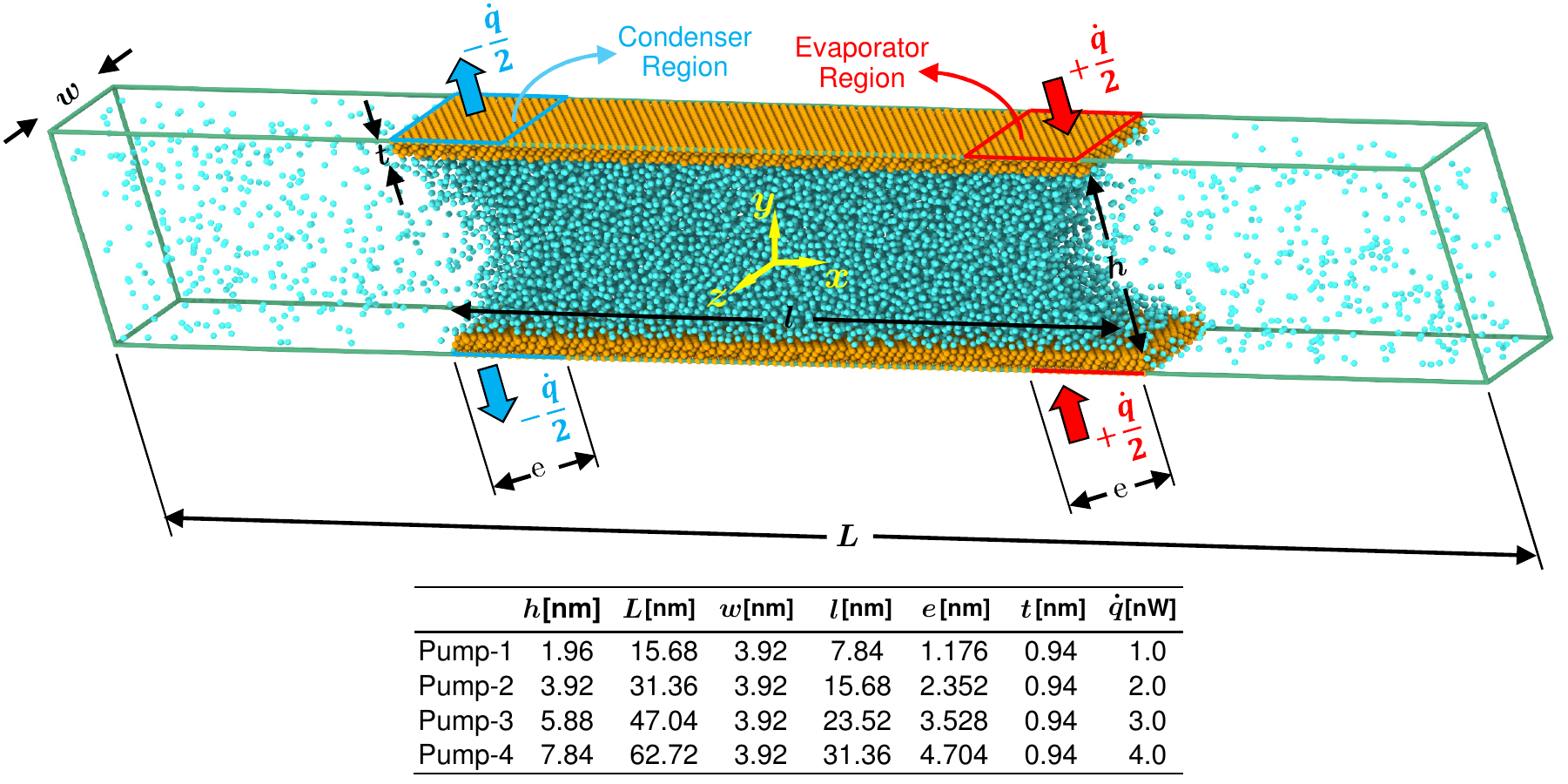,width=6.6in}
	\caption{\label{fig:domain}Schematic of the phase change-driven nanopump system. Liquid phase of the saturated Argon mixture (green spheres) confined between solid Platinum walls (yellow spheres) is pumped from condenser to evaporator by continuous heating and cooling of solid atoms near the ends of the walls. Four different sized (proportionally scaled) nanopumps are simulated to investigate pump performance with scale reduction. Dimensions and heat loads used for each pump are specified in the figure. The wall atoms between evaporator and condenser regions are not allowed to vibrate in order to eliminate heat conduction through the walls.}
\end{figure}

The pump is modelled using molecular dynamics (MD) simulations in a nanochannel system formed by two parallel solid platinum walls. While the size of the simulation domain in the transverse direction ($y$-direction) is determined by the outermost Platinum layers of each wall, simulation domain extends beyond the walls in the longitudinal direction ($x$-direction) as shown in Fig.~\ref{fig:domain}. Inside the channel, saturated Argon mixture condenses to liquid phase due to the interaction of fluid atoms with solid wall atoms, whereas vapor phase of Argon occupies rest of the simulation domain. Initially, formation of stable liquid-gas interfaces, i.e. liquid meniscus structures, is simulated at $110\unit{K}$ in absence of heat loads. Periodic boundary conditions applied in all directions render the system analogous to a liquid block confined between semi-infinite walls within a sufficiently large vapor medium. The amount of fluid atoms is selected such that the condensed phase is always attached to the channel inlets and outlets when the liquid-vapor system equilibrates. To determine the exact thermodynamic state for all simulations, the number of vapor and liquid atoms are counted by \textit{a posteriori} analysis and the total number of Argon atoms is precisely iterated to secure the same mixture quality value, which is $0.06\pm0.005$ for all simulations. Following a $15\unit{ns}$ thermostating and a $15\unit{ns}$ equilibration periods, symmetric and stable liquid-vapor interfaces formed at the ends of the nanochannels. Then, the second part of the simulation is initiated by equally heating and cooling the solid atoms in the heating and cooling zones located at the opposite ends of the nanochannel. Heat transfer to/from liquid is performed by energy injection/extraction from solid atoms instead of thermostat application. This approach eliminates the non-physical temperature jump caused by thermostats \cite{barisik2012}, and preserves the thermodynamic state of the mixture by ensuring zero net heat transfer to the system. Further details about simulations are available in Supplementary Information~1.

\begin{figure}[t]
\centering	
\epsfig{file=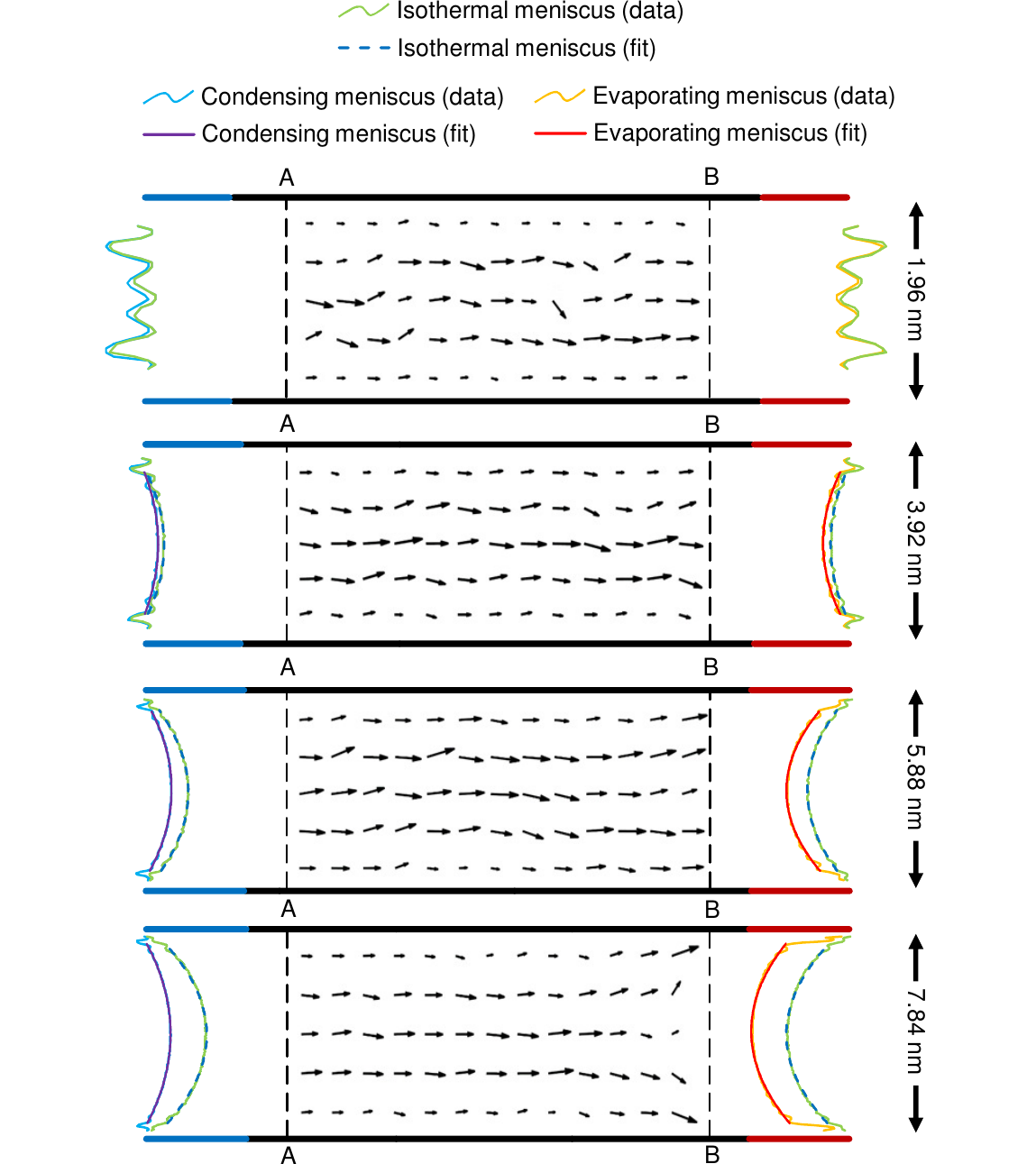,width=4in}
\caption{The resultant liquid-vapor interface profiles and flow fields for the simulated different sized nanopumps. Both isothermal (undeformed) and stedily evaporating/condensing (deformed) interfaces are plotted to demonstrate the interface deformation due to phase change process. Blue and red parts of the channel walls indicate the cooling and heating zones, respectively.  The flow field is presented between sections A-A and B-B by excluding the one-fifth of channel from both ends.}
\label{fig:pumps}
\end{figure}

Scale effects on the working characteristics and efficiency of nanopumps are investigated by simulating four different sized configurations. From the smallest system \mbox{Pump-1} to the largest one \mbox{Pump-4}, the size of the simulation domain and heat loads are increased proportionally as summarized in Fig.~\ref{fig:domain}. The thickness, $t$, and width, $w$, of the solid walls are kept constant to apply identical wall effects on the fluid for each simulation. The resultant liquid-vapor interface profiles and flow fields are demonstrated in Fig.~\ref{fig:pumps}. Although, in all cases, the systems are able to pump the liquid from condenser to evaporator regions, each pump exhibits different liquid-vapor interface profiles. In the absence of phase change, liquid-vapor interfaces are symmetric at both ends, and the profiles of the interfaces are strong function of the channel height. \mbox{Pump-1} does not exhibit meniscus shaped liquid-gas interfaces due to the strong molecular layering of liquid near the walls. With increased channel height, wall force field diminishes sufficiently away from the surfaces, creating bulk flow regions with meniscus shaped liquid-gas interfaces. Unlike their larger scale counterparts, observed meniscus for each case cannot be defined using a single radius of curvature, and a second order polynomial was used to fit the MD data in Fig.~\ref{fig:pumps}. Flatness of the meniscus increases with decreased channel size due to the increase in the relative liquid phase volume. The reason of different liquid/vapor volume ratios for systems at the same thermodynamic state lies in the fact that the average density of the liquid confined within the channel walls reduces with narrower channels due to the prominent near wall density layering \cite{ghorbanian2016}.

Under thermal loading Fig.~\ref{fig:pumps} shows meniscus deformation due to phase change. In the evaporator region, meniscus recedes into the channel and its curvature is increased due to the mass loss and associated evaporation dynamics. On the contrary, meniscus in the condenser region is pushed towards the channel exit, and its profile flattens while it is still attached to channel edges. Phase change associated curvature changes are not observed in \mbox{Pump-1} due to the absence of meniscus. However, interface shift in longitudinal direction is still observable, similar to the other pumps.

\begin{figure}[t!]
	\centering	
	\epsfig{file=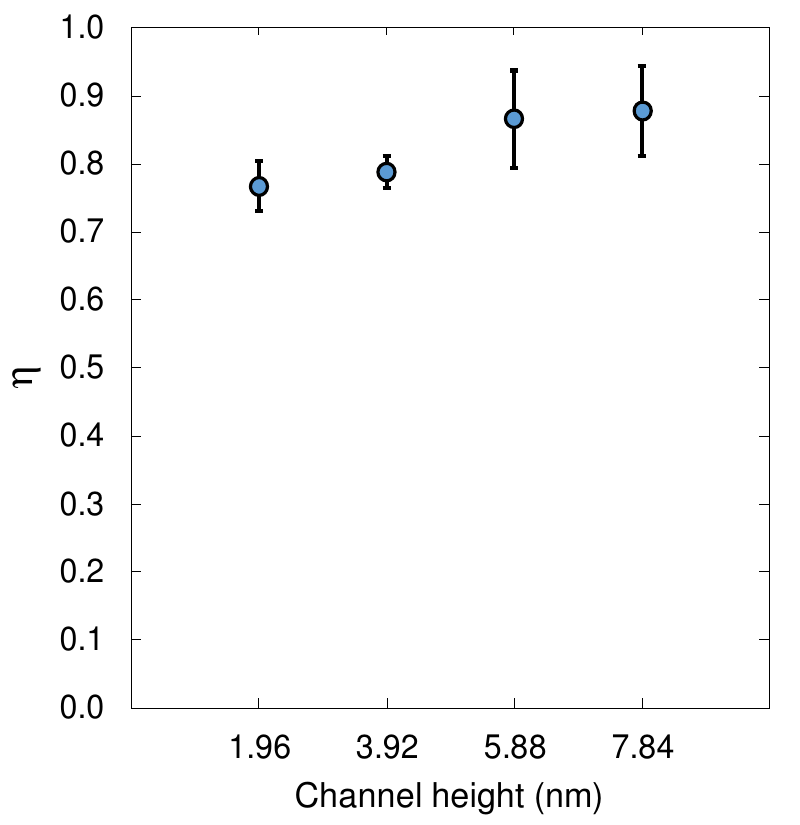,width=2.5in}
	\caption{Non-dimensional liquid mass flow rates as the performance indicators. The average mass flow rate through each pump is normalized by the maximum possible theoretical mass flow rate, which is the ratio of heat input to the enthalpy of evaporation (please see Eqn.(S5) and Eqn.(S6) in Supplementary Information).}
	\label{fig:normalized_mass_flow_rates}
\end{figure}

In the absence of nanoscale effects, continuum theory would suggest an almost identical pumping performance for the geometrically scaled pumps (further discussion is available in Supplementary Information~2) as long as the surface tension forces dominates the gravitational force, i.e. the Bond number is small. However, the scale effects influence the working performance of nanopumps. Efficiency of the phase change driven nanopumps can be determined by their ability to convert the applied heat inputs to the liquid flow. The maximum theoretical mass flow rate through a pump, is equal to the ratio of heat input  to the enthalpy of evaporation, $\dot m_{max}^{therotical} = \dot q_{evap} / h_{fg}$, based on the assumption that all applied heat is used for evaporation. Therefore, the ratio of the average mass flow rate obtained by the simulation to the maximum theoretical mass flow rate, $\eta \equiv \dot m / \dot m_{max}^{therotical}$, can be considered as a good measure of the working performance and these normalized mass flow rates are presented in Fig.~\ref{fig:normalized_mass_flow_rates} for each pump. Including the measurement uncertainties, normalized mass flow rates are estimated between 0.73 and 0.94, which clearly demonstrates the efficient operating capability of phase change driven nanopumps.

The research on the molecular/atomic modeling of phase change has been usually focused on the evaporation from flat nano-thin-film structures \cite{maruyama1999,yi2002,maroo2008,yu2012}. At the same time, the modeling of an evaporating meniscus is cumbersome and rarely investigated using MD. Freund \cite{freund2005} studied steady state evaporation from the menisci of a two-dimensional liquid drop resting on a wall subjected to a symmetric temperature distribution by using spontaneous evaporation and condensation to preserve the shape of the droplet. Maroo and Chung \cite{maroo2010} modeled transient evaporation from a concave meniscus formed by placing the liquid between a lower and an upper platinum wall, with an opening in the upper wall. However, to date, we are unaware of any studies which can construct a stationary and steadily evaporating nanoscale meniscus confined in a capillary conduit at a prescribed thermodynamic state. The unique configuration proposed in this study, on the other hand, allows precise detection of isothermal and phase changing liquid-vapor interfaces as shown in Fig.~\ref{fig:pumps}. From the largest to the smallest pump, two main observations are obvious; (i) the part of the interface approximated by meniscus contracts and eventually disappears at \mbox{Pump-1}, (ii) meniscus deformation due to phase change becomes less prominent. These observations suggest the fact that capillary pressure difference, which sustains the flow in capillary driven systems, diminishes with the decreasing system size and vanishes at \mbox{Pump-1}. However, all pumps, including \mbox{Pump-1}, were able to operate efficiently as demonstrated in Fig.~\ref{fig:normalized_mass_flow_rates}. Therefore, capillary pressure gradient cannot be the sole mechanism responsible for driving the liquid for the proposed nanopumps.
       
\begin{figure}[b!]
	\centering	
	\epsfig{file=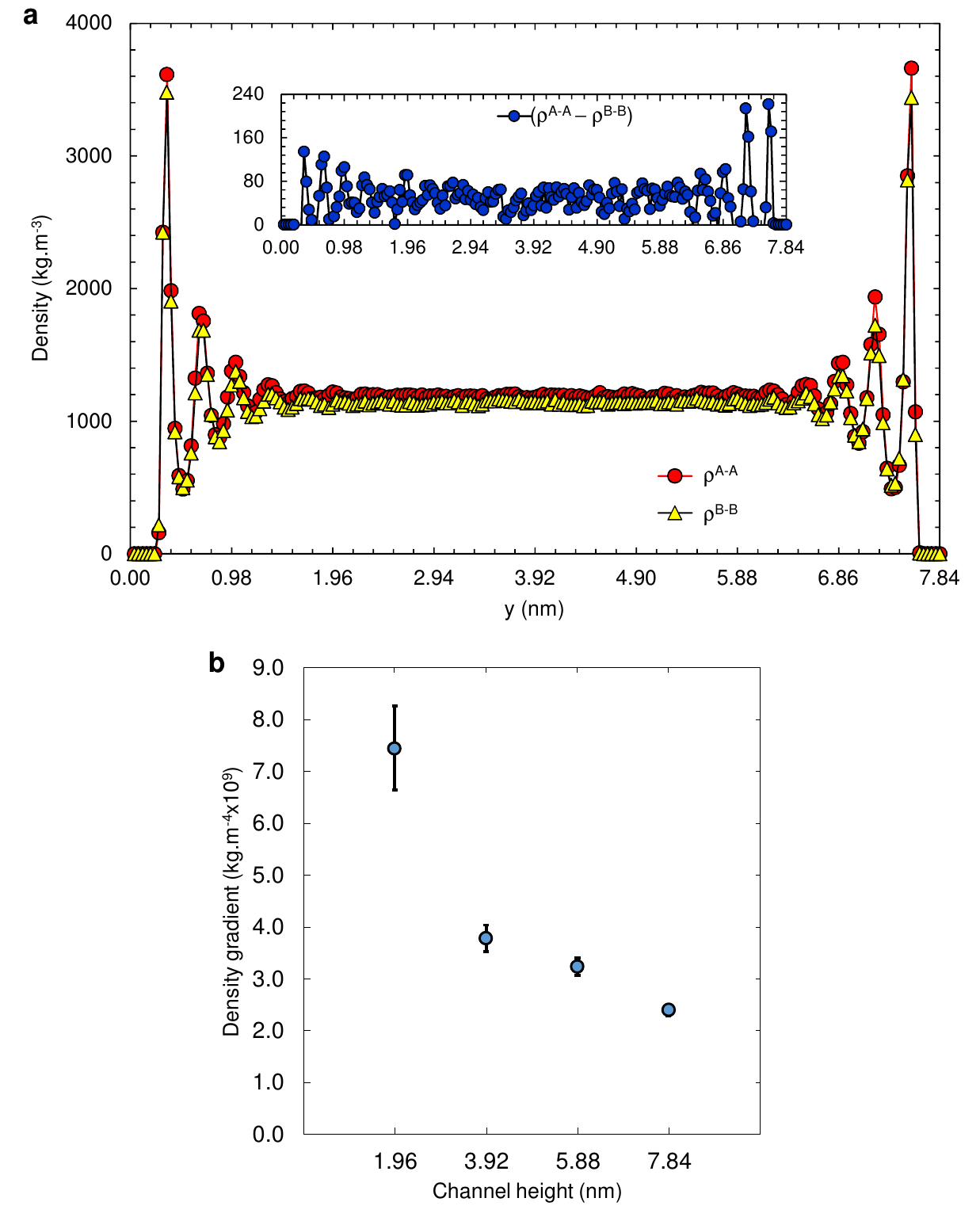,width=4.0in}
	\caption{(a) Density distributions at sections A-A and B-B for \mbox{Pump-4}. The inset shows the difference in density distribution between the two sections. (b) Density gradient of each pump between sections A-A and B-B. Difference between the average densities at these sections are divided to the distance between the sections to estimate the density gradients.}
	\label{fig:AB}
\end{figure}

When density distributions obtained from the simulations are examined (please see Supporting Information~3 for the density distribution of each system), a liquid density gradient between the ends of each pump is noticed. To quantify the density gradient, density distributions at two cross sections, adjacent to the heating and cooling regions, are measured. Fig.~\ref{fig:AB}a shows the density distributions at these cross sections for \mbox{Pump-4}. The inset of Fig.~\ref{fig:AB}a demonstrates that density near the condenser is appreciably higher than the density near the evaporator. Using the average density difference between these cross sections, density gradient of each nanopump is calculated as shown in Fig.~\ref{fig:AB}b. The increase in density gradient with decreasing pump size indicates the fact that diminishing effect of capillary pumping is compensated by a phase change induced mass diffusion. Sharp increase in the density gradient for \mbox{Pump-1} shows that phase change induced molecular scale mass diffusion is the only mechanism driving the liquid in absence of capillary pumping due to the vanished menisci. On the other hand, steadily decreasing density gradient with larger system sizes verifies that for a sufficiently large system, where nanoscale effects are negligible, incompressible liquid flow is driven by capillary pressure gradient as predicted by continuum theory.  

\begin{figure}[b!]
	\centering	
	\epsfig{file=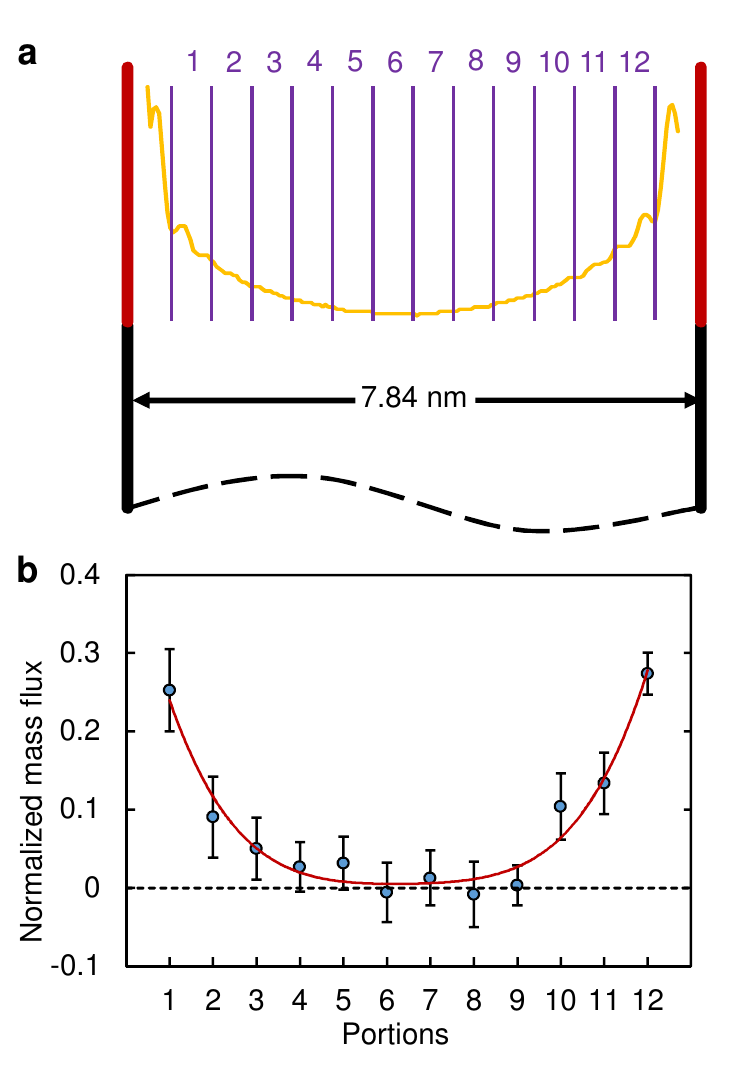,width=3in}
	\caption{Distribution of mass flux at the evaporating interface of \mbox{Pump-4}. (a) Discretizion of liquid-vapor interface. The interface portions closer to the walls than $0.63\unit{nm}$, where interface profile is highly irregular due to severe density fluctuations, are omitted. Remaining part of the interface is divided to 12 portions. (b) Average mass fluxes at each interface portion. A fourth order polynomial fit (red line) is inserted to the data points to guide the eye. In general, evaporation is expected to vanish at the contact line due to the strong intermolecular forces between the solid and liquid atoms. Exclusion of the $0.63\unit{nm}$-thick region adjacent to the walls, on the other hand, prevents the observation of non-evaporating layer at the contact line.}
	\label{fig:evap_mass_flux}
\end{figure}

The existence of different pumping mechanisms can be also inferred from the flow profiles shown in Fig.~\ref{fig:pumps}. As the largest one, \mbox{Pump-4} exhibits the most continuum like behaviour. The flow has almost a parabolic velocity profile suggesting pressure-driven laminar flow of an incompressible Newtonian liquid (i.e. Poiseuille flow). Moreover, the flow is strongly oriented to the contact line near the evaporator, as predicted by the thin film evaporation theory \cite{akkus2017}. To demonstrate the intense evaporation near the contact line, evaporating meniscus of \mbox{Pump-4} is divided to portions as shown in Fig.~\ref{fig:evap_mass_flux}a and evaporative mass flux is calculated at each portion. Distribution of the normalized evaporative mass flux in these regions, shown in Fig.~\ref{fig:evap_mass_flux}b, confirms intense evaporation near the contact line region. The Poiseuille like flow distribution and flow orientation to the contact lines in the evaporator region are also observed for \mbox{Pump-2} and \mbox{Pump-3}. The flow and evaporation characteristics of \mbox{Pump-1}, on the other hand, strongly deviates from the continuum predictions. The liquid molecules experience discrete molecular transport rather than a bulk flow and the evaporation at the interface does not exhibit any regular pattern. These observations are in accordance with the predictions of \cite{travis1997}, which reported the breakdown of continuum flow theory for channel heights smaller than 10 molecular diameters. In our study, the only pump falling into this range is \mbox{Pump-1}.

The pumping mechanism driving the liquid in \mbox{Pump-1} is simply triggered by phase change dynamics. In absence of a meniscus to provide capillary pressure, newly condensed liquid molecules accumulate beneath the condensing interface creating a local higher density region. The fluid-fluid repulsive forces become dominant in this denser region and try to push the atoms away from this zone. Most of the time, atoms cannot overcome the phase change energy barrier to leave the interface and migrate away from the condenser. When these atoms approach to the evaporator end, they gain sufficient energy to leave the liquid phase by continuous heat addition from the evaporator.          

The phase change induced pumping presented in this study differs from the previous thermal gradient driven mechanisms in several aspects. First, temperature gradients, which continuously drive the liquid Argon, were reported in the range of $12.5-81.2\unit{K/nm}$ \cite{liu2010,liu2012,thekkethala2013}. Nonetheless, phase change induced pumping is able to create similar flow rates with only $0.2-0.5\unit{K/nm}$ temperature gradient. Second, thermal creep mechanism which drags the liquid along the positive temperature gradient, was reported to be dominated by a viscous counter flow region, when the hydraulic diameter of the nano-conduit increases \cite{thekkethala2013}. On the contrary, flow direction is always same for phase change driven pumping regardless of the hydraulic diameter. Furthermore, previous studies \cite{liu2009,liu2010,liu2012} reported substantial flow pattern deviations along the channel due to the temperature gradient. However, flow profile is almost unaffected between the evaporator and condenser regions for the phase-change driven nanopumps.         

In summary, we propose a phase change-driven nanopump, which can be precisely controlled by adjusting the energy injection/extraction from solid atoms at the opposite ends of the pump, and demonstrate that this nanopump is able to drive the liquid even at a scale where continuum flow theory breaks down. At this scale, we observed that meniscus structures providing the capillary pressure difference vanish and the flow is driven by a phase change induced molecular mass diffusion mechanism. In absence of meniscus structures, molecular mass diffusion can be still present as in the case of imbibition of liquids in carbon nanotubes. However, the most important exploration of the current study is that phase change induced molecular diffusion mechanism can drive the liquid as efficiently as the capillary pressure induced mechanism. Therefore, these pumps can serve as a part of both NEMS and MEMS devices with similar working efficiencies. Moreover, these pumps can be used for continuous separation of gas mixtures or for nanoscale thermal management applications.

\addcontentsline{toc}{section}{References}
\bibliographystyle{unsrt}

\bibliography{references}

\section*{Acknowledgments}
\addcontentsline{toc}{section}{Acknowledgements}
Authors thank BoHung Kim for helpful discussions. Y.A. acknowledges the financial support of ASELSAN Inc. under scholarship program for postgraduate studies. 


\section*{Author contributions}
\addcontentsline{toc}{section}{Author_contributions}
Y.A. performed molecular dynamics simulations and wrote the manuscript. Both authors contributed most of the ideas and discussed the results. A.B. reviewed and edited the manuscript.

\section*{Competing interests}
\addcontentsline{toc}{section}{Competing_interests}
The authors declare no competing financial interests.

\section*{Additional information}
\textbf{Supplementary information:} (1) Methods; (2) scaling analysis; (3) density distribution of each system.

\end{document}